\documentclass[reviewcopy]{elsart}

\usepackage{amssymb}
\usepackage{graphicx}

\begin{document}

\begin{frontmatter}



\title{Beam-plasma dielectric tensor with \emph{Mathematica}}


\author{A. Bret}
\ead{antoineclaude.bret@uclm.es}

\address{ETSI Industriales, Universidad Castilla-La Mancha, 13071 Ciudad Real, Spain}

\begin{abstract}
We present a \emph{Mathematica} notebook
 allowing for the symbolic
calculation of the $3\times3$ dielectric tensor of a electron-beam
plasma system in the fluid approximation. Calculation is detailed
for a cold relativistic electron beam entering a cold magnetized
plasma, and for arbitrarily oriented wave vectors. We show how one
can elaborate on this example to account for temperatures,
arbitrarily oriented magnetic field or a different kind of
plasma.\\
\bigskip
\textbf{Program summary}\\
\emph{Title of program:} Tensor Mathematica.nb \\
\emph{Catalogue identifier:}   \\
\emph{Program summary URL:}  \\
\emph{Program obtainable from:} CPC Program Library, Queen
University of Belfast, N.
 Ireland\\
\emph{Computer for which the program is designed and others on
which it
has been tested: Computers:} Any computer running \emph{Mathematica} 4.1. Tested on DELL Dimension 5100 and IBM ThinkPad T42.\\
\emph{Installations:} ETSI Industriales, Universidad Castilla la Mancha, Ciudad Real, Spain\\
\emph{Operating systems under which the program has been tested:} Windows XP Pro\\
\emph{Programming language used:} \emph{Mathematica} 4.1
\\
\emph{Memory required to execute with typical data:} 7.17 Mbytes\\
\emph{No. of lines in distributed program, including test data,
etc.:} 19 \\
\emph{No. of bytes in distributed program, including test data,
etc.:} 4 172 \\
\emph{Distribution format:} .nb \\
\emph{Nature of physical problem:} The dielectric tensor of a
relativistic beam plasma system may be quite involved to calculate
symbolically when considering a magnetized plasma, kinetic
pressure, collisions between species, and so on. The present
\emph{Mathematica} Notebook performs the symbolic computation
in terms of some usual dimensionless variables. \\
\emph{Method of solution:} The linearized relativistic fluid
equations are directly entered and solved by \emph{Mathematica} to
express the first order expression
 of the current. This expression is then introduced into a combination of Faraday and Amp\`{e}re Maxwell's equations to give the dielectric tensor. Some additional manipulations are needed to express the
  result in terms of the dimensionless variables.\\
\emph{Restrictions on the complexity of the problem:} Temperature
effects are limited to small, i.e non-relativistic, temperatures.
The kinetic counterpart of the present \emph{Mathematica} Notebook cannot be implemented because \emph{Mathematica} will usually not compute the required integrals.\\
\emph{Typical running time:} About 1 minute on a Intel Centrino
1.5 Ghz Laptop with 512 Mo of RAM. \\
\emph{Unusual features of the
program:} none.

\end{abstract}


\end{frontmatter}


\section{Introduction}
The calculation of the dielectric tensor of a beam plasma system
is a recurrent problem in plasma physics. Many efforts have been
dedicated recently to such issue because of the Fast Ignition
Scenario for inertial thermonuclear fusion \cite{Tabak,Tabak2005}.
According to this view, the Deuterium Tritium target is first
compressed by means of some driver. Then, the compressed fuel is
ignited by a relativistic electron beam generated by a petawatt
laser shot. Such scenario implies therefore the interaction of a
relativistic electron beam with a plasma. This kind of
interaction, and its magnetized counterpart, is also relevant to
astrophysics, in particular when investigating the relativistic
jets of microquasars \cite{fender}, active galactic nuclei
\cite{zensus}, gamma ray burst production scenarios
\cite{Piran2004} or pulsar winds \cite{gallant}. Theoretical works
on these subjects are usually focused on the instabilities of the
system. Although many of them demands a kinetic treatment to be
fully described, the fluid equations can set some very relevant
guidelines, especially when the system is not too hot.
Furthermore, it has been known for long that in the relativistic
regime, instabilities with arbitrarily orientated wave vectors may
be essential
\cite{fainberg,Godfrey1975,Califano1,Califano2,Califano3}. One can
therefore figure out how some refined kinetic theory may lead to
almost unsolvable calculations whereas the fluid formalism is
still tractable. For example, a detailed description of the
collisional filamentation instability ($\mathbf{k}\perp$ beam)
including the movement of the background ions plasma, and
accounting for temperatures, was first performed through the fluid
equations \cite{Honda}. The very same equations were used to
explore the growth rate of unstable modes with arbitrarily
oriented wave vectors (with respect to the beam) when a
relativistic electron beam enters a plasma
\cite{fainberg,Califano1,Califano2,Califano3}. The results were
found crucial as it was demonstrated that the fastest growing
modes were indeed found for obliquely propagating waves, and the
kinetic counterpart of these models has only been considered very
recently \cite{BretPRE2004,BretPRE2005,BretPRL2005}. As far as the
magnetized case is concerned, the kinetic formalism has been
thoroughly investigated for wave vectors parallel and normal to
the beam \cite{Cary1981,Tautz2005,Tautz2006}. But
 the unstable oblique modes, which once again turn to be the most
 unstable in many cases, could only be explored through the fluid
 formalism \cite{Godfrey1975}.

 It has been demonstrated that the fluid
 equations yield the same first order temperature corrections
 than the kinetic theory for oblique modes, and the roles of both
 beam and plasma parallel and perpendicular temperatures are
 retrieved \cite{BretPoPFluide}. The fluid approximation is thus definitely a tool of
 paramount importance to deal with beam plasma instabilities.
 Additionally, it generally yields a polynomial dispersion
 equation for which numerical resolution is immediate.
 Nevertheless, even the fluid tensor can be analytically involved
 when considering arbitrarily oriented wave vectors, a guiding
 magnetic field, temperatures, and so on \cite{BretPoPMagnet}. Indeed, on can think about
 any model based on whether the system is relativistic or
 not, collisional or not, magnetized or not, hot or cold\ldots
  Most of these models have not been implemented yet, and each one
 should leave a quite complicated dielectric tensor.

 This is why
 a \emph{Mathematica} notebook has been developed which allows for the
 symbolic calculation of the fluid tensor, once the parameters of
 the system have been set. The basic system we study here is a
 cold relativistic electron beam entering a cold magnetized plasma with return current.
 As the reader shall check, the notebook is very easy to adapt the
 different scenarios (ion beam, temperatures, pair plasma...). The paper is
 structured as follow: we start introducing the theory leading to
 the fluid dielectric tensor in section \ref{sec:theory}. The
 \emph{Mathematica} notebook is then explained step by step in section
 \ref{sec:notebook}, and we show how it can be modified to include temperatures or collisions before the comments and conclusion section.

\section{\label{sec:theory}Theory}
We consider a beam of density $n_b$, velocity $\mathbf{V}_b$ and
relativistic factor $\gamma_b=1/(1-V_b^2/c^2)$ entering a plasma
of density $n_p$. Ions from the plasma are considered as a fixed
neutralizing background, and an electron plasma return current
flows at velocity $\mathbf{V}_p$ such as
$n_p\mathbf{V}_p=n_b\mathbf{V}_b$. The system is thus charge and
current neutralized. We do not make any assumptions on the ratio
$n_b/n_p$ so that the return current can turn relativistic for
beam densities approaching, or even equalling, the plasma one. We
set the $z$ axis along the beam velocity and align the static
magnetic field along this very axis. The wave vector investigated
lies in the $(x,z)$ plan without loss of generality
\cite{Godfrey1975}, and we define the angle $\theta$ between
$\mathbf{k}$ and $\mathbf{V}_b\parallel \mathbf{B}_0\parallel z$
through $k_z=k\cos\theta$ and $k_x=k\sin\theta$. The dielectric
tensor of the system is obtained starting with the fluid equations
for each species $j=p$ for plasma electrons and $j=b$ for the beam
ones,
\begin{equation}\label{eq:conservation}
  \frac{\partial n_j}{\partial t}-\nabla\cdot (n_j\mathbf{v}_j) =
  0,
\end{equation}
\begin{equation}\label{eq:force}
 \frac{\partial \mathbf{p}_j}{\partial t}+(\mathbf{v}_j\cdot\nabla) \mathbf{p}_j =
 q\left(\mathbf{E}+\frac{\mathbf{v}_j\times \mathbf{B}}{c}\right),
\end{equation}
where $\mathbf{p}_j=\gamma_j m\mathbf{v}_j$, $m$ the electron mass
and $q<0$ its charge. The equations are then linearized according
to a standard procedure \cite{Godfrey1975}, assuming small
variations of the variables according to $\exp(i\mathbf{k}\cdot
\mathbf{r}-i\omega t)$. With the subscripts 0 and 1 denoting the
equilibrium and perturbed quantities respectively, the linearized
conservation equation (\ref{eq:conservation}) yields
\begin{equation}\label{eq:conservationL}
  n_{j1} = n_{j0} \frac{\mathbf{k}\cdot \mathbf{v}_{j1}}{\omega -\mathbf{k}\cdot
  \mathbf{v}_{j0}},
\end{equation}
and the force equation (\ref{eq:force}) gives,
\begin{eqnarray}\label{eq:forceL}
  &&i m \gamma_j (\mathbf{k}\cdot
\mathbf{v}_{j0}-\omega)\left(\mathbf{v}_{j1}+\frac{\gamma_j^2}{c^2}(\mathbf{v}_{j0}\cdot
\mathbf{v}_{j1})\mathbf{v}_{j0}\right)\nonumber\\
   &=& q\left(\mathbf{E}_{1}+\frac{(\mathbf{v}_{j0}+\mathbf{v}_{j1})\times
   \mathbf{B}_0+\mathbf{v}_{j0}\times
   \mathbf{B}_1}{c}\right),
\end{eqnarray}
where $i^2=-1$. Through Maxwell-Faraday equations, the field
$\mathbf{B}_1$ is then replaced by
$(c/\omega)\mathbf{k}\times\mathbf{E}_1$ so that the perturbed
velocities $\mathbf{v}_{j1}$ can be explained in terms of
$\mathbf{E}_1$ resolving the tensorial equations
(\ref{eq:forceL}). Once the velocities are obtained, the perturbed
densities can be expressed in terms of the electric field using
Eqs. (\ref{eq:conservationL}). Finally, the linear expression of
the current is found in terms of $\mathbf{E}_1$ through,
\begin{equation}\label{eq:current}
   \mathbf{J} = q\sum_{j=p,b}
   n_{j0}\mathbf{v}_{j1}+n_{j1}\mathbf{v}_{j0},
\end{equation}
and the system is closed combining Maxwell Faraday and Maxwell
Amp\`{e}re equations,
\begin{equation}\label{eq:Maxwell}
  \frac{c^2}{\omega^2}\mathbf{k}\times(\mathbf{k}\times \mathbf{E_1})+\mathbf{E_1} + \frac{4
  i
  \pi}{\omega}\mathbf{J} = 0.
\end{equation}
Inserting the current expression from Eq. (\ref{eq:current}) into
Eq. (\ref{eq:Maxwell}) yields an equation of the kind
$\mathcal{T}(\mathbf{E_1})=0$, and the dispersion equation reads
det$\mathcal{T}=0$.

The Mathematica notebook we describe in the next section performs
a symbolic computation of the tensor $\mathcal{T}$ and the
dispersion equation det$\mathcal{T}=0$, in terms of the usual
\cite{Ichimaru} reduced variables of the problem
\begin{equation}\label{eq:param}
\mathbf{Z}=\frac{\mathbf{k}V_b}{\omega_p},~~x=
   \frac{\omega}{\omega_p},~~\alpha=\frac{n_b}{n_p},~~\beta=\frac{V_b}{c},~~\Omega_B=\frac{\omega_b}{\omega_p},
\end{equation}
where $\omega_p^2=4\pi n_p q^2/m$ is the electron plasma frequency
and $\omega_b=|q|B_0/mc$ the electron cyclotron frequency.

\section{\label{sec:notebook}\emph{Mathematica} implementation}
For the most part, \emph{Mathematica} is used to solve the
tensorial equations (\ref{eq:forceL}) for $\mathbf{v}_{j1}$ and
extract the tensors $\mathcal{T}$ from Eqs.
(\ref{eq:current},\ref{eq:Maxwell}). We start declaring the
variables corresponding to the wave vector, the  electric field,
the beam and plasma drift velocities and the  magnetic field,

\emph{In[1]:= }\textbf{k = \{kx, 0, kz\}; E1 = \{E1x, E1y, E1z\};
V0b = \{0, 0, Vb\}; V0p = \{0, 0, Vp\}; B0=\{0, 0, $\mathbf{m~ c~
\omega b/q}$\}; B1 = c Cross[k, E1]/$\mathbf{\omega}$; vb1 =
\{vb1x, vb1y, vb1z\}; vp1 = \{vp1x, vp1y, vp1z\};}

Note that Maxwell Faraday's equation is already implemented in the
definition of $\mathbf{B1}$. The wave vector has no component
along the $y$ axis and the beam and plasma drift velocities only
have one along the $z$ axis. The guiding magnetic field is set
along $z$ and  defined in terms of the cyclotron frequency
$\omega_b$. This will be useful later when introducing the
dimensionless parameters (\ref{eq:param}).

We then have \emph{Mathematica} solve Eqs. (\ref{eq:forceL}) for
the beam and the plasma. The left hand side of the equation is not
as simple as in the non-relativistic case because the $\gamma$
factors of the beam and the plasma modify the linearization
procedure. We write this part of the equations in a tensorial form
in \emph{Mathematica} defining the tensors \textbf{Mp} and
\textbf{Mb} such as ``left hand
side''=\textbf{Mj}$^{-1}$.\textbf{vj1} with,

\emph{In[2]:= }\textbf{Mb=\{\{$\frac{\mathbf{i}}{\gamma
b(\omega-kz Vb)}$,0,0\},\{0,$\frac{\mathbf{i}}{\gamma b(\omega-kz
Vb)}$,0\},\{0,0,$\frac{\mathbf{i}}{\gamma b^3(\omega-kz
Vb)}$\}\}};

\emph{In[3]:= }\textbf{Mp=\{\{$\frac{\mathbf{i}}{\gamma
p(\omega-kz Vp)}$,0,0\},\{0,$\frac{\mathbf{i}}{\gamma p(\omega-kz
Vp)}$,0\},\{0,0,$\frac{\mathbf{i}}{\gamma p^3(\omega-kz
Vp)}$\}\}};

where $\mathbf{i}^2=-1$. The reader will notice that relativistic
effects are more pronounced in the beam direction due to the
$\gamma^3$ factors in the $zz$ component. We now have
\emph{Mathematica} solve the tensorial Eqs. (\ref{eq:forceL}). For
better clarity, we first define them

\emph{In[4]:=}\textbf{EqVb=vb1-Dot[Mb,$\frac{\mathbf{q}}{\mathbf{m}}\left(\mathbf{E1}+\frac{\mathrm{Cross[\mathbf{V0b+vb1,B0}]}}{\mathbf{c}}\right)$]-Dot[Mb,$\frac{\mathbf{q}}{\mathbf{m}}\left(\frac{\mathrm{Cross[\mathbf{V0b,B1}]}}{\mathrm{\mathbf{c}}}\right)$];}

\emph{In[5]:=}\textbf{EqVp=vp1-Dot[Mp,$\frac{\mathbf{q}}{\mathbf{m}}\left(\mathbf{E1}+\frac{\mathrm{Cross[\mathbf{V0p+vp1,B0}]}}{\mathbf{c}}\right)$]-Dot[Mp,$\frac{\mathbf{q}}{\mathbf{m}}\left(\frac{\mathrm{Cross[\mathbf{V0p,B1}]}}{\mathrm{\mathbf{c}}}\right)$];}

before we solve them,

\emph{In[6]:=}\textbf{Vb1=FullSimplify[vb1/.
Solve[EqVb==0,vb1][[1]]];}

\emph{In[7]:=}\textbf{Vp1=FullSimplify[vp1/.
Solve[EqVp==0,vp1][[1]]];}

Note that the \textbf{Vb}'s, with capital ``V'', store the
solutions of the equations whereas the \textbf{vb}'s are the
variables. This is why the \textbf{Vb}'s do not need to be defined
at the beginning (see \emph{In[1]}) of the notebook; they are
implicitly defined here.

Now that we have the values of the perturbed velocities, we can
derive the perturbed densities from Eqs. (\ref{eq:conservationL}),

\emph{In[8]:=}\textbf{Nb1=FullSimplify[$\mathbf{\omega
pb}^2\frac{\mathbf{m}}{4\pi
\mathbf{q}^2}\frac{\mathbf{Dot[k,Vb1]}}{\mathbf{\omega-Dot[k,V0b]}}$];}

\emph{In[9]:=}\textbf{Np1=FullSimplify[$\mathbf{\omega
pp}^2\frac{\mathbf{m}}{4\pi
\mathbf{q}^2}\frac{\mathbf{Dot[k,Vp1]}}{\mathbf{\omega-Dot[k,V0p]}}$];}

Here again, we prepare the introduction of the reduced variables
(\ref{eq:param}) by expressing the equilibrium beam and plasma
electronic densities in terms of the beam and plasma electronic
frequencies.

We can now have \emph{Mathematica} calculate the current according
to Eq. (\ref{eq:current}),

\emph{In[10]:=}

\textbf{J=FullSimplify[q$\left(\mathbf{\omega
pp}^2\frac{\mathbf{m}}{4\pi
\mathbf{q}^2}\mathbf{Vp1}+\mathbf{\omega
pb}^2\frac{\mathbf{m}}{4\pi \mathbf{q}^2}\mathbf{Vb1} \mathbf{+
Np1 V0p+Nb1 V0b}\right)$];}

We now have the symbolic expression of the current \textbf{J}. In
order to find the tensor $\mathcal{T}$ yielding the dispersion
equation, we need to explain first the current tensor. This is
performed through,

\emph{In[11]:=}\textbf{M=}
\begin{displaymath}
\left(
\begin{array}{lll}
  \mathbf{Coefficient[J[[1]],E1x]} & \mathbf{Coefficient[J[[1]],E1y]} & \mathbf{Coefficient[J[[1]],E1z]} \\
  \mathbf{Coefficient[J[[2]],E1x]} & \mathbf{Coefficient[J[[2]],E1y]} & \mathbf{Coefficient[J[[2]],E1z]} \\
  \mathbf{Coefficient[J[[3]],E1x]} & \mathbf{Coefficient[J[[3]],E1y]} & \mathbf{Coefficient[J[[3]],E1z]} \\
\end{array}
\right)\mathbf{;}
\end{displaymath}

which just extract the tensor elements from the expression of
\textbf{J}. We now turn to Eq. (\ref{eq:Maxwell}) where we explain
the tensor elements of the quantity
$c^2\mathbf{k}\times(\mathbf{k}\times
\mathbf{E_1})+\omega^2\mathbf{E_1}$,

\emph{In[12]:=}\textbf{M0=$\mathbf{c}^2$
Cross[k,Cross[k,E1]]+$\omega^2$E1 ;}

\emph{In[13]:=}\textbf{M1=}
\begin{displaymath}
\left(
\begin{array}{lll}
  \mathbf{Coefficient[M0[[1]],E1x]} & \mathbf{Coefficient[M0[[1]],E1y]} & \mathbf{Coefficient[M0[[1]],E1z]} \\
  \mathbf{Coefficient[M0[[2]],E1x]} & \mathbf{Coefficient[M0[[2]],E1y]} & \mathbf{Coefficient[M0[[2]],E1z]} \\
  \mathbf{Coefficient[M0[[3]],E1x]} & \mathbf{Coefficient[M0[[3]],E1y]} & \mathbf{Coefficient[M0[[3]],E1z]} \\
\end{array}
\right)\textbf{;}
\end{displaymath}

We can finally express the tensor $\mathcal{T}$ defined by
$\mathcal{T}(\mathbf{E})$=0 as

\emph{In[14]:=}\textbf{T=M1+4 i $\pi ~\omega$ M;}

At this stage of the notebook, we could take the determinant of
the tensor to obtain the dispersion equation. Let us first
introduce the dimensionless variables (\ref{eq:param}) through,

\emph{In[15]:=}\textbf{T=T /. \{Vp $\rightarrow -\alpha$ Vb, kz
$\rightarrow \omega \mathbf{pp}$ Zz/Vb, kx $\rightarrow \omega
\mathbf{pp}$ Zx/Vb, $\omega \mathbf{pb}^2\rightarrow \alpha
~\omega \mathbf{pp}^2$, $\omega  \rightarrow \mathbf{x} ~\omega
\mathbf{pp}$, $\omega \mathbf{b} \rightarrow  \Omega \mathbf{b}
~\omega\mathbf{pp}$\}};

and,

\emph{In[16]:=}\textbf{T=T /. \{Vb $\rightarrow \beta$ c\}}

\emph{Mathematica} leaves here some $\omega \mathbf{pp}$'s which
should simplify between each others. It is enough to perform

\emph{In[17]:=}\textbf{T=T /. \{$\omega\mathbf{pp} \rightarrow
1$\};}

and a simple

\emph{In[18]:=}\textbf{MatrixForm[FullSimplify[T]]}

displays the result. The dispersion equation of the system is
eventually obtained through

\emph{In[19]:=}\textbf{DisperEq=Det[T]}

The notebook evaluation takes 1 minute on a Laptop running a 1.5
GHz Pentium Centrino under Windows XP Pro. This delay can be
shortened down to 10 seconds by suppressing all the
\textbf{FullSimplify} routines while leaving a
\textbf{Simplify[T]} in entry \emph{18}, but the final result is
much less concise and readable.

\section{Comments and Conclusion}
In this paper, we have described a \emph{Mathematica} notebook
performing the symbolic evaluation of the dielectric tensor of a
beam plasma system. Starting from the linearized fluid equations,
the notebook expresses the dielectric tensor, and eventually the
dispersion equation, is terms of some usual dimensionless
parameters. This notebook has been so far applied to the treatment
of the temperature dependant non magnetized and magnetized
problems (see Refs \cite{BretPoPFluide,BretPoPMagnet}). Indeed,
the procedure is very easy to adapt to different settings.

When including beam or plasma temperatures, one  adds a pressure
term $-\nabla P_j/n_j$ on the right hand side of the force
equations (\ref{eq:force}). Setting then $\nabla P_i = 3 k_BT_i
\nabla n_i$ \cite{Honda,Kruer} if dealing only with electron
motion, one only needs to add to the notebook entries \emph{4} and
\emph{5} the terms (\textbf{i}$^2$=-1)

\textbf{-3i Tj k $\frac{\mathbf{Dot[k,
vj1]}}{\omega-\mathbf{Dot[k, V0j]}}$},

where \textbf{j=p} for the plasma, and \textbf{b} for the beam.
When considering anisotropic temperatures \cite{BretPoPFluide},
one just needs to define a temperature tensor \textbf{Tj} for each
species \textbf{j}, and replace the scalar product \textbf{Tj k}
by the tensorial one \textbf{Dot[Tj,k]} in both entries. Of
course, a correct treatment of electromagnetic instabilities
generally requires a kinetic formalism instead of a fluid one.
However, kinetic calculations cannot be systematically entrusted
to \emph{Mathematica}, as is the case here. The reason why is that
the relativistic factors $\gamma$ encountered in the kinetic
quadratures are coupling the integrations along the three
components of the momentum. According to the distribution
functions considered, the quadratures may be calculable through
some ad hoc change of variables, if they can be calculated at all.
At any rate, the process cannot be systematized enough for
\emph{Mathematica} to handle it.

As far as the magnetic field is concerned, its direction can be
changed from entry \emph{1} without any modification of the next
entries. When dealing with the motion of ions, or even with one of
these pair plasmas involved in the pulsar problems
\cite{GedalinPRL}, one just need to modify the conservation and
force equations according to the properties of the species
investigated. It is even possible to add more equations to account
for more species because the resolution involves only the force
and the conservation equations of one single specie at a time
before the perturbed quantities merge together in entry \emph{10}
to compute the current \textbf{J}.

The notebook can thus be easily adapted to different settings and
allows for a quick symbolic calculation of the dielectric tensor
and the dispersion equation, even for an elaborated fluid model.

\section{Acknowledgements}
This work has been  achieved under projects FTN 2003-00721 of the
Spanish Ministerio de Educaci\'{o}n y Ciencia and PAI-05-045 of
the Consejer\'{i}a de Educaci\'{o}n y Ciencia de la Junta de
Comunidades de Castilla-La Mancha.

\end{document}